\documentclass[
twocolumn,
preprintnumbers,
nofootinbib,
prd,aps
]{revtex4}

\usepackage{epsf}
\usepackage{latexsym}
\usepackage{amssymb}
\usepackage{epsf}
\usepackage{amsmath}
\usepackage{slashed}

\usepackage{graphicx}
\usepackage{comment}

\allowdisplaybreaks[1]

\begin{document}


\newcommand{\arctanh}{{\,\rm arctanh}}

\newcommand{\rem}[1]{{\bf #1}}

\renewcommand{\topfraction}{0.8}

\preprint{UT-19-15}

\title{

  Bounce Configuration from Gradient Flow


}

\author{
So Chigusa$^{(a)}$, Takeo Moroi$^{(a)}$ and Yutaro Shoji$^{(b)}$
}

\affiliation{
$^{(a)}$Department of Physics, The University of Tokyo, Tokyo 113-0033, Japan
\\
$^{(b)}$Kobayashi-Maskawa Institute for the Origin
of Particles and the Universe, Nagoya University, Nagoya 464-8602, Japan
}

\date{June, 2019}

\begin{abstract}

  Based on the gradient flow, we propose a new method to determine the
  bounce configuration for false vacuum decay.  Our method is
  applicable to a large class of models with multiple fields.  Since
  the bounce is a saddle point of an action, a naive gradient flow
  method which minimizes the action does not work.  We point out that
  a simple modification of the flow equation can make the bounce its
  stable fixed point while the false vacuum configuration an unstable
  one.  Consequently, the bounce configuration can be obtained simply
  by following the flow without a careful choice of an initial
  configuration.  With numerical analysis, we confirm the validity of
  our claim, checking that the flow equation we propose indeed has
  solutions that flow into the bounce configuration.

\end{abstract}

\maketitle

\renewcommand{\thefootnote}{\#\arabic{footnote}}

\noindent\underline{\it Introduction:} Study of false vacua (and
metastable states) has been important in various fields, like particle
physics, cosmology, nuclear physics, condensed matter physics, and so
on.  For example, in the field of particle physics and cosmology, the
stability of the electroweak vacuum has been attracted much attention.
In particular, taking the best-fit values of the observed top-quark
and Higgs-boson masses, and assuming that the standard model of
particle physics is valid up to a very high scale (like the Planck
scale), the electroweak vacuum is metastable \cite{Isidori:2001bm,
  Degrassi:2012ry}. It is because the Higgs quartic coupling constant
becomes negative at a high energy scale due to the renormalization
group effects.  Performing precise calculation based on relativistic
quantum field theory, the decay rate of the electroweak vacuum per
unit volume is known to be $\sim 10^{-582}\ {\rm Gyr^{-1}Gpc^{-3}}$
\cite{Chigusa:2017dux, Chigusa:2018uuj, Andreassen:2017rzq}, with
which the stability of our universe looks plausible for the present
cosmic time scale.  However, this conclusion may be altered with the
introduction of new physics beyond the standard model. The studies of
the stability of the electroweak vacuum in such new physics models
remain important.

In relativistic quantum field theory, the decay of the false vacuum is
mainly induced by the field configuration called ``bounce''
\cite{Coleman:1977py, Callan:1977pt, Coleman:aspectsof}.  Bounce is a
configuration obeying the classical equation of motion (EOM) derived
from the Euclidean action.  With the bounce, which we denote as
$\bar\phi$, the decay rate of the false vacuum per unit volume is given
in the following form:
\begin{align}
  \gamma = {\cal A} e^{-{\cal S}[\bar{\phi}]},
\end{align}
where ${\cal S}[\bar{\phi}]$ is the bounce action while ${\cal A}$ is a
prefactor.  The decay rate is highly sensitive to the bounce action so
that the profile of the bounce should be well understood for an accurate
calculation of the decay rate.

In spite of the importance of determining the bounce configuration
with a generic potential, it is not easy in general. The difficulty
mainly comes from the fact that the bounce is a saddle point, not a
minimum, of the action.  Consequently, the fluctuation matrix around
the bounce has a negative eigenvalue and a small fluctuation
destabilizes the bounce.  Although there have been many attempts to
find methods to determine the bounce configuration with overcoming
this difficulty \cite{Claudson:1983et, Kusenko:1995jv, Kusenko:1996jn,
  Dasgupta:1996qu, Moreno:1998bq, John:1998ip, Cline:1998rc,
  Cline:1999wi, Konstandin:2006nd, Wainwright:2011kj, Akula:2016gpl,
  Masoumi:2016wot, Espinosa:2018hue, Jinno:2018dek, Espinosa:2018szu,
  Athron:2019nbd}, new ideas are still awaited for a detailed
understanding of the bounce as well as for a precise calculation of
its configuration via numerical analysis.

In this letter, we propose a new method to determine the bounce
configuration, where we use the gradient flow method.\footnote
{Refs.\ \cite{Cline:1998rc,Cline:1999wi} also discuss the possibility to
  use gradient flow to derive the bounce configuration.  The idea of
  Refs.\ \cite{Cline:1998rc,Cline:1999wi} is to introduce back steps
  during the flow, and is different from ours.}
It does not work if we naively use the action, ${\cal S}$, to
calculate the gradient in the configuration space.  The failure of the
naive method is due to the negative eigenvalue mode as we have
mentioned.  We discuss that, with a simple modification of the flow
equation, the bounce configuration can become a stable fixed point
while it makes the false vacuum and other stable solutions of the
classical EOM unstable.\footnote
{There is another approach to make the negative eigenvalue mode
  harmless by adding new terms, which vanish with the bounce
  configuration, to the action \cite{Kusenko:1995jv}.  In this
  approach, the bounce becomes a minimum of the improved action. It
  is, however, not guaranteed that the obtained configuration is
  indeed the bounce.}
We also show that, with numerical analysis, the bounce configuration
can be obtained by solving the flow equation we propose.

\vspace{2mm}
\noindent\underline{\it Formulation:}
We adopt the action in the following form:
\begin{align}
  {\cal S}[\phi] =
  \int d^D x
  \left[
    \frac{1}{2}
    \partial_\mu \phi_A \partial_\mu \phi_A + V(\phi)
    \right],
\end{align}
where $D$ is the dimension of the Euclidean space, $\phi_A$ denotes a
real scalar field (with $A$ being flavor index), and $V$ is the scalar
potential.\footnote
{Here and hereafter, the summation over the repeated flavor
indices is implicit.}
Because the bounce is a spherical object \cite{Coleman:1977th,
  Blum:2016ipp}, $\phi_A$ depends only on the radial coordinate of the
  Euclidean space, $r$, and obeys the following classical EOM:
\begin{align}
  \left.
    \frac{\delta {\cal S} [\phi]}{\delta \phi_A}
  \right|_{\phi\rightarrow\bar{\phi}}
  =
  - \partial_r^2 \bar{\phi}_A - \frac{D-1}{r} \partial_r \bar{\phi}_A
  + \left. \frac{\partial V}{\partial \phi_A} \right|_{\phi\rightarrow\bar{\phi}}
  = 0,
\end{align}
satisfying the following boundary conditions:
\begin{align}
  \partial_r \bar{\phi}_A (r=0) = 0, ~~~
  \bar{\phi}_A(r\rightarrow\infty) = v_A,
  \label{BounceBCs}
\end{align}
where $v_A$ is the field amplitude of the $A$-th scalar field at the
false vacuum.  In the following, we explain how we obtain the bounce
configuration.

Before going into the details, we introduce the fluctuation operator, which
plays important roles in our discussion.  In $D$-dimensional Euclidean
space, the fluctuation operator around the bounce is given by
\begin{align}
  {\cal M}_{AB}
  \equiv
  - \left( \partial_r^2 + \frac{D-1}{r} \partial_r \right) \delta_{AB}
  + \left. \frac{\partial^2 V}{\partial \phi_A \partial \phi_B}
  \right|_{\phi\rightarrow\bar{\phi}}.
  \label{M0}
\end{align}
Spherical fluctuations around the bounce can be expressed as linear
combinations of the eigenfunctions of ${\cal M}_{AB}$.  We denote
eigenfunctions of ${\cal M}_{AB}$ as $\chi_{n,A}$ ($n=-1$, $1$, $2$,
$\cdots$), {\it i.e.},
\begin{align}
  {\cal M}_{AB} \chi_{n,B} = \lambda_n \chi_{n,A},
\end{align}
where $\lambda_n$ is the eigenvalue.  The eigenfunctions should
satisfy the following boundary conditions:
\begin{align}
  \partial_r \chi_{n,A} (r=0) = 0,~~~
  \chi_{n,A} (r\rightarrow\infty) = 0,
  \label{BCs}
\end{align}
and are normalized as
\begin{align}
  \langle \chi_n | \chi_{n'} \rangle = \delta_{nn'},
\end{align}
where the inner product of two sets of functions is defined as
\begin{align}
  \langle f | f' \rangle \equiv
  \int_0^\infty dr r^{D-1} f_A (r) f'_A (r).
\end{align}
An important property of the bounce is that the fluctuation operator
around the bounce has one negative eigenvalue \cite{Callan:1977pt},
which we call $\lambda_{-1}<0$.  We also assume that all the other
eigenvalues are positive.

Hereafter, we discuss a method in which a function $\Phi_A (r,s)$, with
$s$ being the ``flow time,'' evolves to the bounce configuration as
$s\rightarrow\infty$.  The initial profile $\Phi_A (r,s=0)$ is required
to satisfy the boundary conditions same as the bounce (see
\eqref{BounceBCs}). Then, the boundary conditions are kept during the
flow by the flow equation introduced below.  In other words, $\Phi_A
(r,s)-\bar{\phi}_A$ stays in the configuration space spanned by the
eigenfunctions of ${\cal M}_{AB}$, and hence $\Phi_A (r,s)$ can be
expressed as
\begin{align}
  \Phi_A (r,s) = \bar{\phi}_A (r) + \sum_n a_n (s) \chi_{n,A} (r).
  \label{phiexp}
\end{align}

To obtain the bounce configuration using a flow equation, we need to
flip the sign of the negative eigenvalue around the bounce. The flow
equation we propose is as follows:\footnote
{We adopted Eq.\ \eqref{floweq} as our flow equation.  Another
  possibility may be $\partial_s \Phi_A = {\cal M}_{AB}F_B$; with only
  the terms linear in $a_n$ being kept, it gives $\dot{a}_n=-\lambda_n^2
  a_n$, and hence all the fluctuations around the solution of the
  classical EOM damp.  We leave its study as a future project.}
\begin{align}
  \partial_s \Phi_A (r,s) = F_A (r,s)
  - \beta \langle F | g \rangle
  g_A(r),
  \label{floweq}
\end{align}
where $\beta$ is a dimensionless constant.  Here, $g_A$ is a function
satisfying the same boundary conditions as $\chi_{n,A}$ (see Eq.\
\eqref{BCs}), and is normalized as
\begin{align}
  \langle g | g \rangle = 1.
\end{align}
We expand $g_A$ as
\begin{align}
  g_A (r) = \sum_n c_n \chi_{n,A} (r),
\end{align}
where $\sum_n c_n^2=1$.  In addition,
\begin{align}
  F_A \equiv - \frac{\delta {\cal S} [\Phi]}{\delta \Phi_A}
  = \partial_r^2 \Phi_A + \frac{D-1}{r} \partial_r \Phi_A
  - \frac{\partial V (\Phi)}{\partial \Phi_A}.
\end{align}
Notice that $F_A$ satisfies the same boundary conditions as
$\chi_{n,A}$ as far as $\Phi_A$ is in the form of Eq.\ \eqref{phiexp},
guaranteeing that $\Phi_A$ is expressed as in Eq.\ \eqref{phiexp} for
any value of $s$.  As we will see in the following, the second term in
the right-hand side of Eq.\ \eqref{floweq} can make the bounce
configuration a stable fixed point.

Importantly, for $\beta\neq 1$, any fixed point solution of Eq.\
\eqref{floweq}, which satisfies $\partial_s\Phi_A=0$, is a solution of
the classical EOM, $F_A=0$.  This can be understood by using the
following relation:
\begin{align}
  \langle\partial_s \Phi|g\rangle=(1-\beta)\langle F|g \rangle.
  \label{<Phidot|g>}
\end{align}
If $\partial_s\Phi_A=0$ is realized with non-vanishing $F_A$, $F_A$ and
$g_A$ should be proportional to each other to satisfy the flow equation.
Such a requirement contradicts with Eq.\ \eqref{<Phidot|g>} because the
left-hand side of Eq.\ \eqref{<Phidot|g>} vanishes while the right-hand
side does not.  It also implies that the flow equation of our proposal
does not have any unwanted fixed point that does not correspond to a
solution of the classical EOM.

Now, we show that, with properly choosing $\beta$ and $g_A$, the flow
equation \eqref{floweq} has solutions that evolve to the bounce as
$s\rightarrow\infty$.  For this purpose, we analyze the flow around the
bounce configuration, where $a_n$ can be treated as a small quantity.
Keeping terms linear in $a_n$, we obtain
\begin{align}
  F_A \simeq
  - {\cal M}_{AB} (\Phi_B - \bar{\phi}_B)
  = - \sum_n \lambda_n a_n \chi_{n,A},
\end{align}
which gives
$\langle F | g \rangle \simeq- \sum_n \lambda_n c_n a_n$.
Thus, the flow equation gives
\begin{align}
  \dot{a}_n \simeq
  -\lambda_n a_n + \beta
  \sum_m c_n c_m \lambda_m a_m
  \equiv
  - \sum_m \Gamma_{nm} (\beta) a_m,
  \label{adot}
\end{align}
where the ``dot'' is the derivative with respect to $s$.

If we consider the naive flow equation which minimizes the action
({\it i.e.} the case with $\beta=0$), we obtain $\dot{a}_n \simeq
-\lambda_n a_n$. Then, the coefficient of the mode with the negative
eigenvalue $\lambda_{-1}$ grows with flow time.  This is the reason
why the naive gradient flow method does not work to find the bounce.

With a non-vanishing value of $\beta$, the above conclusion may
change.  To see this, we express $\Gamma (\beta)$ in Eq.\ \eqref{adot}
in the matrix form:
\begin{align}
  \Gamma (\beta) =
  \left(
    {\rm I} - \beta \vec{c}\, \vec{c}^{\,T}
  \right) \mbox{diag} (\lambda_{-1}, \lambda_1, \lambda_2, \cdots),
\end{align}
where ${\rm I}$ is the unit matrix, the superscript, $T$, denotes the
transpose, and
\begin{align}
  \vec{c} =
  (c_{-1}, c_1, c_2, \cdots)^{\,T}.
\end{align}
If the real parts of all the eigenvalues of $\Gamma$ are positive,
$\Phi_A (r,s\rightarrow\infty)=\bar{\phi}_A(r)$ is realized.  Notice
that $\vec{c}$ is an eigenvector of the matrix $({\rm I} - \beta
\vec{c}\, \vec{c}^{\,T})$ with the eigenvalue of $(1-\beta)$.  In
addition, all the other eigenvalues are $1$ because $({\rm I} -\beta
\vec{c}\, \vec{c}^{\,T})\vec{v}_\perp=\vec{v}_\perp$ if
$\vec{c}^{\,T}\vec{v}_\perp=0$.  Thus, we have
\begin{align}
  \mbox{det}
  \left(
    {\rm I} - \beta \vec{c}\, \vec{c}^{\,T}
  \right)
  = 1-\beta,
\end{align}
and hence
\begin{align}
  \mbox{det} \Gamma (\beta) =
  ( 1-\beta ) \prod_n \lambda_n.
\end{align}
For $\beta>1$, $\mbox{det} \Gamma>0$ (because $\prod_n\lambda_n<0$),
which opens a possibility to make the real parts of all the
eigenvalues of $\Gamma(\beta>1)$ positive.

The evolution of $\vec{a}\equiv(a_{-1},a_1,a_2,\cdots)^T$ is
complicated in general because $\Gamma$ is not symmetric.  If we
consider the condition $\partial_s(\vec{a}^{\,T}\vec{a})<0$, which is
a sufficient condition for the bounce to be a stable fixed point, the
discussion becomes simpler; it requires $\Gamma+\Gamma^T$ to be
positive definite.  In order for $\Gamma(\beta>1)+\Gamma^T(\beta>1)$
to be positive definite,
$\vec{c}^{\,T}[\Gamma(\beta>1)+\Gamma^T(\beta>1)]\vec{c}>0$ should
hold, and hence
\begin{align}
  \langle g | {\cal M}g \rangle = \sum_n \lambda_n c_n^2 < 0,
  \label{Condition_g}
\end{align}
which gives a guideline in choosing the function,
$g_A$.\footnote
{As noted, $\partial_s(\vec{a}^{\,T}\vec{a})<0$ is a sufficient
  condition and, with numerical analysis, we found that the bounce may
  become a stable fixed point with $g_A$ which does not satisfy
  $\langle g|\mathcal M g\rangle<0$. }
In addition, because the smallest eigenvalue of $\Gamma+\Gamma^T$ is
smaller than its smallest diagonal element, $\beta$ should be in the
following range:
\begin{align}
  \frac{1}{c_{-1}^2} <
  \beta <
  \frac{1}{\mbox{max}_{n\geq 1} c_n^2}.
\end{align}
We can see that a choice of $g_A$ with larger $|c_{-1}|$ and smaller
$|c_{n\geq 1}|$ is better though their exact values are unknown until the
bounce is obtained. We will discuss the choice of $g_A$ later.

Let us summarize the properties of the flow equation of
Eq.~\eqref{floweq}.  If $\Phi_A (r,s\rightarrow\infty)$ converges with
$\beta>1$, it is guaranteed that (i) $\Phi_A (r,s\rightarrow\infty)$
is a solution of the classical EOM, (ii) $\Phi_A
(r,s\rightarrow\infty)$ satisfies the boundary conditions relevant for
the bounce, and (iii) the fluctuation operator around $\Phi_A
(r,s\rightarrow\infty)$ has one negative eigenvalue.  The statement
(iii) is due to the fact that the real parts of all the eigenvalues of
$\Gamma(\beta>1)$ should be positive to make $\Phi_A
(r,s\rightarrow\infty)$ stable against fluctuations, which implies
that $\Gamma(\beta=0)$ has one negative eigenvalue assuming that there
is no degeneracy in the eigenvalues of $\Gamma(\beta=1)$.  Thus,
$\Phi_A (r,s\rightarrow\infty)$ obtained with $\beta>1$ is expected to
be the bounce configuration.  We also emphasize that, because of
(iii), all the stable fixed points for $\Gamma(\beta=0)$ are
destabilized. Thus, for example, the resultant configuration $\Phi_A
(r,s\rightarrow\infty)$ cannot be the false vacuum configuration when
$\beta>1$.

\vspace{2mm}
\noindent\underline{\it Numerical Analysis:} So far, we have studied the
behavior of the fluctuations around the bounce and have seen that the
bounce configuration can become a fixed point of the flow equation.  In
the following, by using numerical calculations, we explicitly show that
there exist solutions that indeed flow to the bounce configuration.

To perform numerical calculations, the function $g_A$ should be fixed.
We take
\begin{align}
  g_A \propto r \partial_r \Phi_A,
\end{align}
which is based on the following consideration. Since we have
\begin{align}
  &
  \int_0^\infty dr r^{D-1}
  (r \partial_r \bar{\phi}_A) {\cal M}_{AB} (r \partial_r \bar{\phi}_B)
  \nonumber \\ &
  = - (D-2) \int_0^\infty dr r^{D-1}
  (\partial_r \bar{\phi}_A)(\partial_r \bar{\phi}_A),
\end{align}
the condition \eqref{Condition_g} is satisfied for $D>2$. It implies
that $g_A$ of our choice has a large $|c_{-1}|$. In addition, it
satisfies the relevant boundary conditions.

To solve Eq.~\eqref{floweq}, we discretize the radius coordinate, $r$,
and solve the ordinary differential equations with respect to $s$.  To
impose the boundary conditions of Eq.~\eqref{BounceBCs}, it is better to
shrink $r\in(0,\infty)$ into $x\in(0,1)$ with
\begin{equation}
 x=\tanh\left(\frac{r}{R}\right),
\end{equation}
and attach the endpoints, $x=0$ and $x=1$.  In our analysis, $R$ is
taken to the size of the bounce.  Then, we discretize $x$ into $N+1$
lattice points:
\begin{equation}
 x_0=0,~x_1=\frac{1}{N},~x_2=\frac{2}{N},\cdots,~x_N=1.
\end{equation}
The flow equation in terms of $x$ is given by
\begin{align}
  \partial_\sigma \hat \Phi_A (x,\sigma) = 2\sigma(1-x^2)\left[
 \hat F_A (x,\sigma)
  - \beta
  \frac{\langle \hat F | \hat g \rangle_x}
  {\langle \hat g | \hat g \rangle_x}
 \hat g_A(x,\sigma)
 \right],
  \label{floweq_numerical}
\end{align}
where $\sigma=s^{1/2}/R$,
\begin{align}
 \hat \Phi_A(x,\sigma)&=R^{\frac{D}{2}-1}\Phi_A(R\arctanh(x),R^2\sigma^2),\\
 \hat g_A(x,\sigma)&=\arctanh(x)\partial_x\hat \Phi_A(x,\sigma),
\end{align}
and
\begin{align}
  \hat F_A= &\, (1-x^2)\partial_x^2\hat \Phi_A
  +\frac{D-1-2x\arctanh(x)}{\arctanh(x)}\partial_x\hat \Phi_A
  \nonumber \\ &\,
  -\frac{R^{\frac{D}{2}+1}}{1-x^2}\frac{\partial V}{\partial\phi_A}.
\end{align}
Furthermore, the inner product is defined as
\begin{equation}
 \langle \hat f|\hat f'\rangle_x\equiv
 \int_0^1dx(1-x^2)\arctanh^{D-1}(x)\hat f_A(x)\hat f'_A(x).
\end{equation}
Here, we adopt the second-order central differences for the derivatives
with respect to $x$.

In numerically solving the flow equation, we take the following
initial configuration of $\hat\Phi_A$:
\begin{equation}
 \hat\Phi_A(x_n,0)=R^{\frac{D}{2}-1}[w_A+x_n^2(v_A-w_A)],
\end{equation}
where $w_A=\Phi_A(x_0,0)$ is a constant; $\Phi_A(x_0,0)$ is set to be
somewhere near the true vacuum (see figures).  At each step of the
flow, $\hat\Phi_A(x_1,\sigma),\cdots,\hat\Phi_A(x_{N-1},\sigma)$ are
determined by Eq.~\eqref{floweq_numerical}, while the endpoint values
are fixed by the boundary conditions:
\begin{align}
  &\hat\Phi_A(x_0,\sigma)=
  \frac{4\hat\Phi_A(x_1,\sigma)-\hat\Phi_A(x_2,\sigma)}{3},
  \label{BC_origin}
  \\
  &\hat\Phi_A(x_N,\sigma)=R^{\frac{D}{2}-1}v_A.
  \label{BC_infty}
\end{align}
Notice that Eq.\ \eqref{BC_origin} is equivalent to $\partial_r\Phi_A=0$
at $r=0$ (where we used $\partial_x\hat\Phi_A (x_0,\sigma)\simeq
\partial_x\hat\Phi_A(x_1,\sigma)-\partial_x^2\hat\Phi_A(x_1,\sigma)/N$),
while Eq.\ \eqref{BC_infty} guarantees $\Phi_A(r=\infty)=v_A$.  After
the convergence of $\hat\Phi_A$, we calculate the bounce action.  With
the use of the equation of motion for the bounce, the action is
calculated as
\begin{equation}
  {\cal S}[\bar{\phi}]=\frac{S_{D-1}}{D}
  \langle\partial_x\hat\Phi|\partial_x\hat\Phi\rangle_x,
\end{equation}
where $S_{D-1}$ is the surface area of the $(D-1)$-dimensional sphere.

Here, we take $D=3$, and use the following benchmark potentials for
single- and double-scalar cases:
\begin{align}
  V^{\rm (single)} &=\frac{1}{4}\phi^4-\frac{k_1+1}{3}\phi^3
  +\frac{k_1}{2}\phi^2,\\
  V^{\rm (double)} &=\left(\phi_x^2+5\phi_y^2\right)
  \left[5(\phi_x-1)^2+(\phi_y-1)^2\right]\nonumber\\
  &\hspace{3ex}+k_2\left(\frac{1}{4}\phi_y^4-\frac{1}{3}\phi_y^3\right),
\end{align}
with $k_1$ and $k_2$ being constants.  With our choices of parameters,
the false and true vacua of $V^{\rm (single)}$ ($V^{\rm (double)}$)
are $\phi=0$ and $1$ ($(\phi_x,\phi_y)=(0,0)$ and $(1,1)$),
respectively.  For the single-scalar (double-scalar) case, we take
$k_1=0.47$ and $0.2$ ($k_2=2$ and $80$), which correspond to the
thin-walled and thick-walled bounces, respectively.

We emphasize that the initial configurations, the model parameters,
$\beta$, and the other lattice parameters are not special choices. We
checked that the bounce can be obtained as a result of flow with
generic choices of parameters.

\begin{figure}[t]
  \begin{minipage}{0.95\linewidth}
    \includegraphics[width=\linewidth]{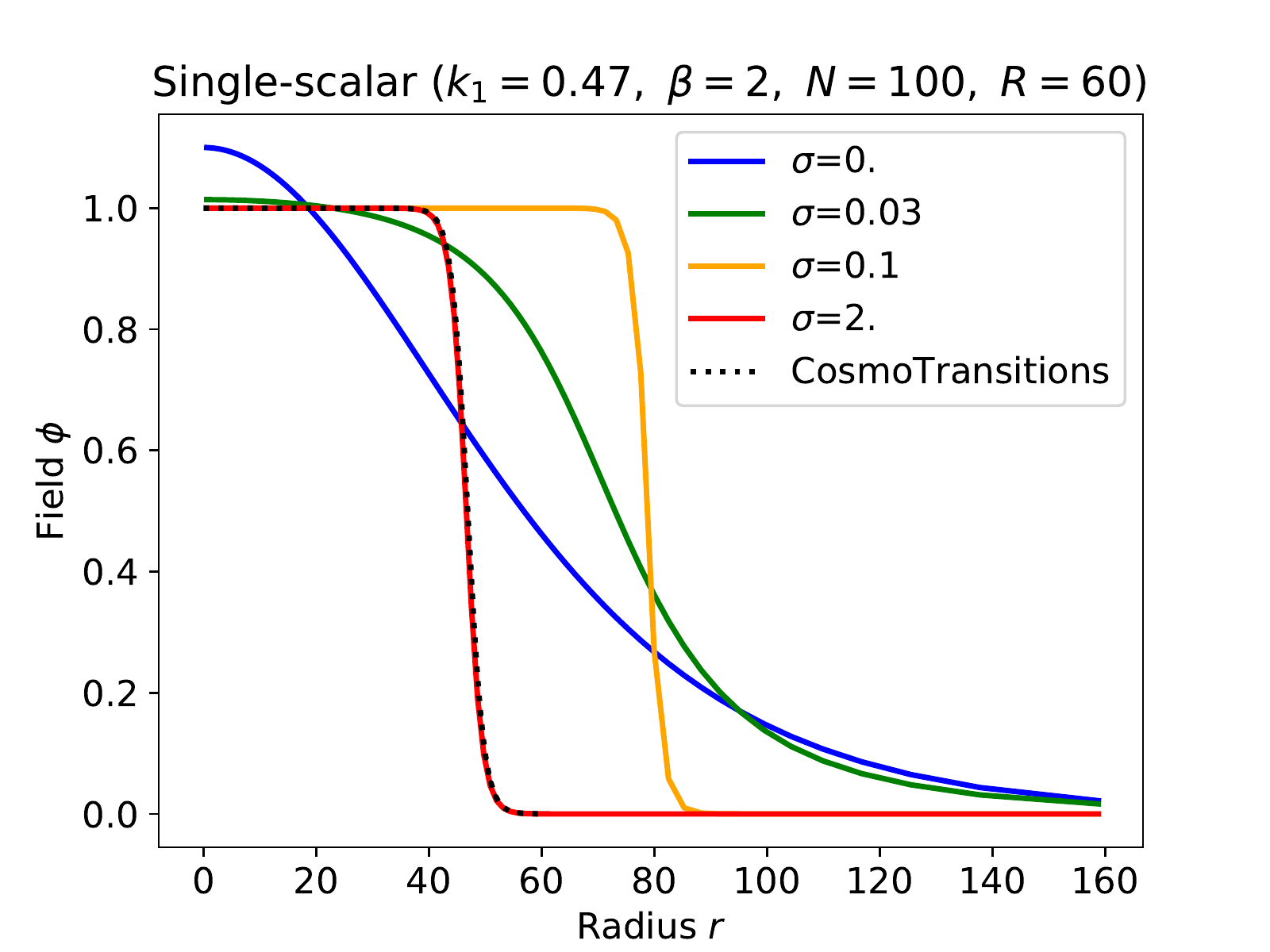}
  \end{minipage}
  \begin{minipage}{0.95\linewidth}
    \includegraphics[width=\linewidth]{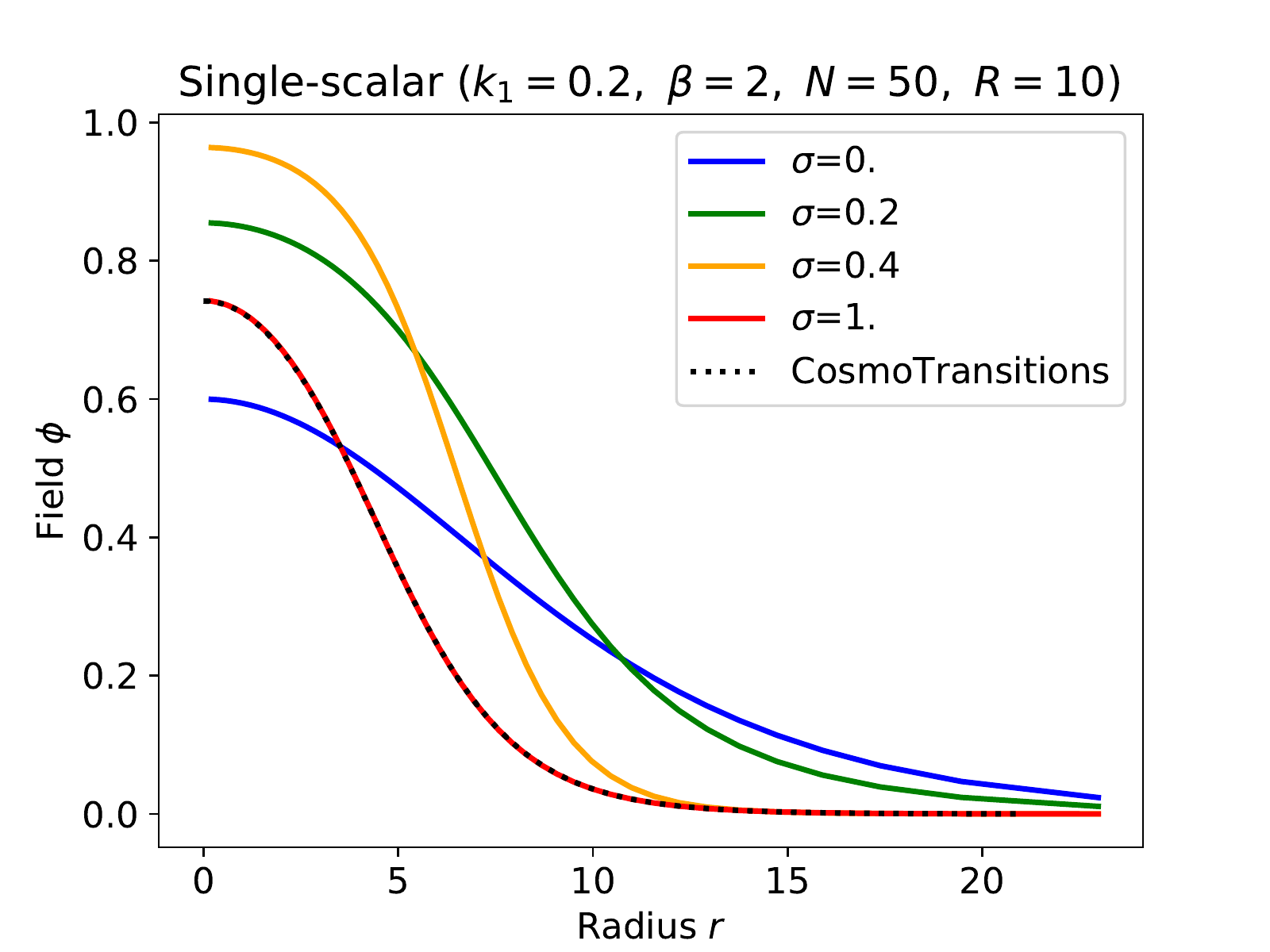}
  \end{minipage}
  \caption{The flow of $\Phi$ for the single-scalar potential.  The
    configuration is shown with the solid line for each $\sigma$. The
    results of {\tt CosmoTransitions} are also shown with dotted
    lines. The parameters are shown on the top of each panel.}
 \label{fig_flow1}
\end{figure}

\begin{figure}[t]
  \begin{minipage}{0.95\linewidth}
    \includegraphics[width=\linewidth]{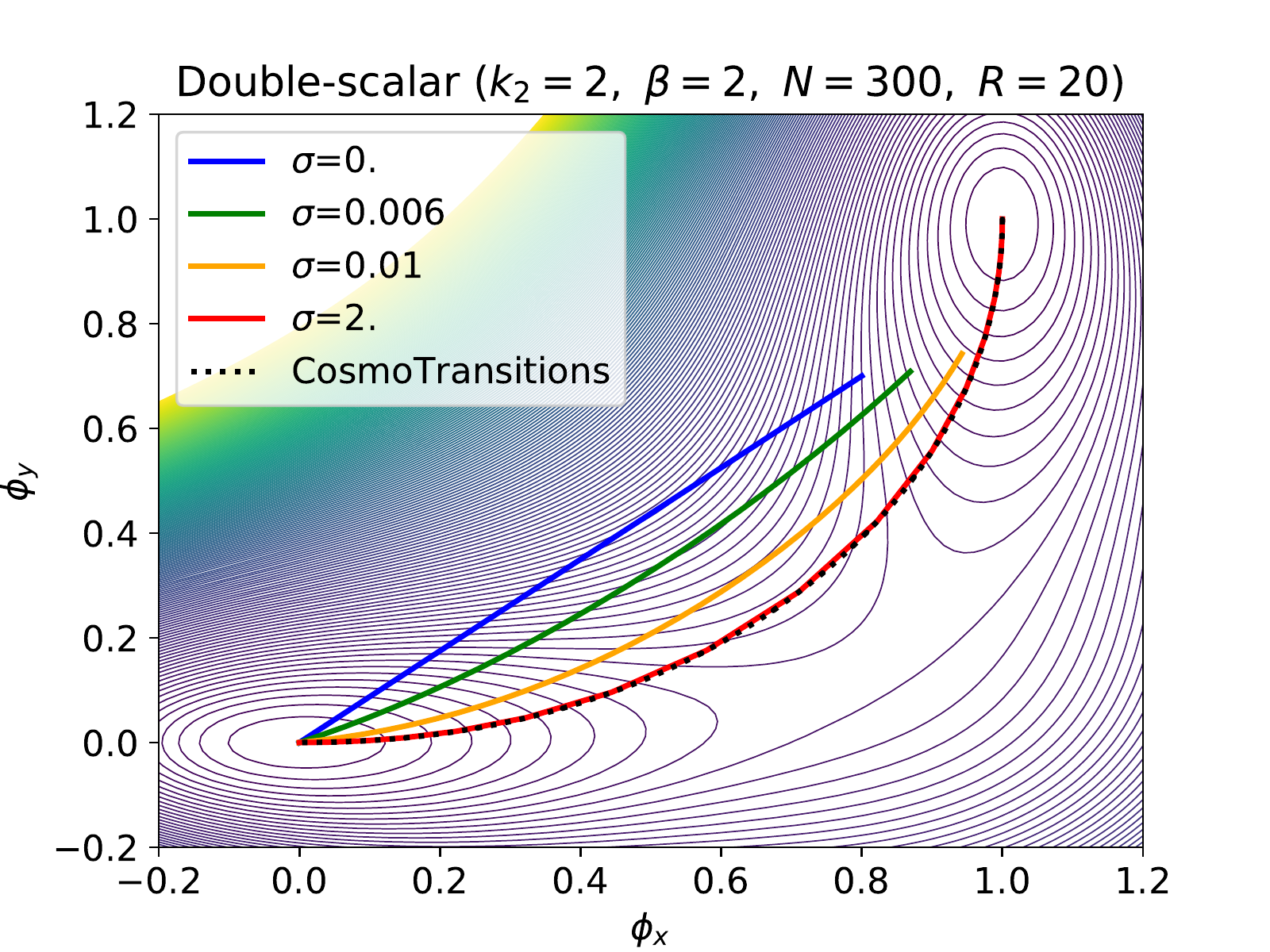}
  \end{minipage}
  \begin{minipage}{0.95\linewidth}
    \includegraphics[width=\linewidth]{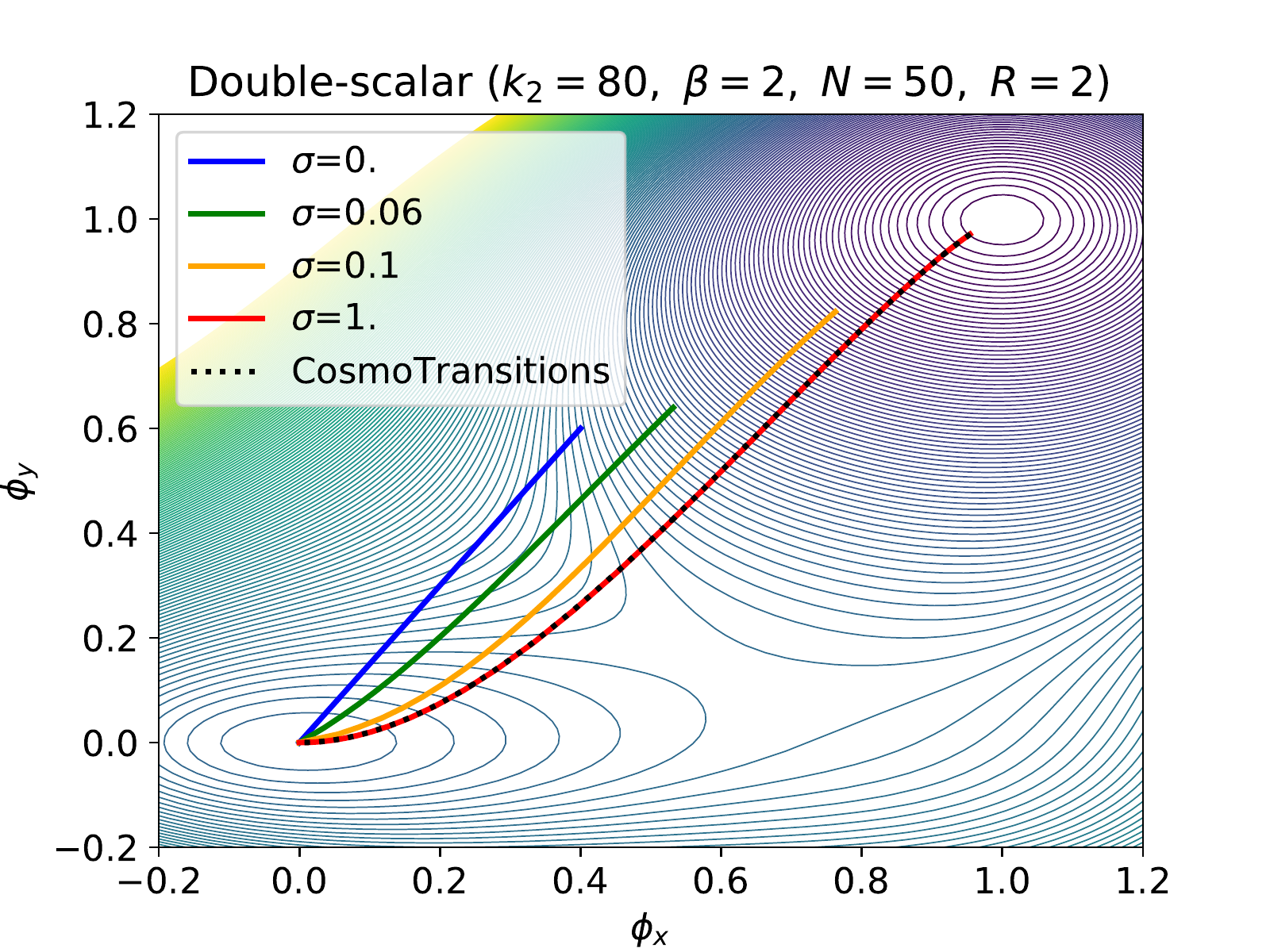}
  \end{minipage}
  \caption{The flow of $\Phi$ for the double-scalar potential.  The
    configuration is shown with the solid line for each $\sigma$. The
    results of {\tt CosmoTransitions} are also shown with dotted
    lines. The parameters are shown on the top of each panel.  We also
    show the contours of constant $V^{\rm (double)}$. }
 \label{fig_flow2}
\end{figure}

The flows of $\Phi_A$ based on our method are shown in
Figs.~\ref{fig_flow1} (for the single-scalar potential) and
\ref{fig_flow2} (for the double-scalar potential) with the solid
lines.  In each figure, the solid line with the largest $\sigma$ shows
the field configuration after the convergence; we checked that the
flow after such an epoch is negligible.  For comparison, we also
determine the bounces for the same models by using {\tt
  CosmoTransitions} \cite{Wainwright:2011kj},
which are shown with the dotted lines.
(We use the default control parameters of {\tt CosmoTransitions}
  except for {\tt fRatioConv=0.001}, which improves the accuracy in
  the multi-field calculation.)
In addition, we calculate the
bounce action ${\cal S}[\bar{\phi}]$ for each case:
\begin{align}
 \mathcal S^{\rm (single)}_{k_1=0.47}[\bar{\phi}]&=
 \begin{cases}
  1086.6,&{\rm Our~result},\\
  1092.8,&{\tt CosmoTransitions},
 \end{cases}\\
 \mathcal S^{\rm (single)}_{k_1=0.2}[\bar{\phi}]&=
 \begin{cases}
  6.6360,&{\rm Our~result},\\
  6.6490,&{\tt CosmoTransitions},
 \end{cases}\\
 \mathcal S^{\rm (double)}_{k_2=2}[\bar{\phi}]&=
 \begin{cases}
  1763.7,&{\rm Our~result},\\
  1767.7,&{\tt CosmoTransitions},
 \end{cases}\\
 \mathcal S^{\rm (double)}_{k_2=80}[\bar{\phi}]&=
 \begin{cases}
  4.4585,&{\rm Our~result},\\
  4.4661,&{\tt CosmoTransitions}.
 \end{cases}
\end{align}
As we can see from the figures and the values of $\mathcal
S[\bar{\phi}]$, our results well agree with those of {\tt
  CosmoTransitions}.  This strongly suggests the validity of our
gradient flow method to determine the bounce configuration.

\vspace{2mm}
\noindent\underline{\it Summary:} In this letter, we have proposed a
new method to determine the bounce configuration.  We have pointed out
that the bounce configuration can be a stable solution of the flow
equation given in Eq.\ \eqref{floweq}.  If the solution of the flow
equation evolves to a fixed configuration for $\beta>1$, the resultant
configuration is always a saddle point of the action, {\it i.e.}, the
bounce configuration.  We have analytically shown that the negative
eigenvalue mode, which destabilizes the bounce configuration, can be
made harmless and that the bounce configuration can become a stable
solution of the flow equation.  We have verified our claims by
numerical analysis.  We believe that our method of finding the bounce
configuration is simple, powerful, and useful in many cases.  It can
be used in multi-scalar cases and can easily be implemented to
numerical code. We also comment that, even though we have concentrated
on the bounce, our method is applicable to a generic problem to find a
saddle point in a configuration space.

\vspace{2mm}
\noindent\underline{\it Acknowledgements:}
This work was supported by JSPS KAKENHI Grant (Nos.\ 17J00813 [SC],
16H06490 [TM], 18K03608 [TM], and 16H06492 [YS]).


\begin{thebibliography}{99}

\vspace{-1.7cm}

\bibitem{Isidori:2001bm}
  G.~Isidori, G.~Ridolfi and A.~Strumia,
  Nucl.\ Phys.\ B {\bf 609} (2001) 387
  [hep-ph/0104016].

\bibitem{Degrassi:2012ry}
  G.~Degrassi, S.~Di Vita, J.~Elias-Miro, J.~R.~Espinosa,
  G.~F.~Giudice, G.~Isidori and A.~Strumia,
  JHEP {\bf 1208} (2012) 098
  [arXiv:1205.6497 [hep-ph]].

\bibitem{Chigusa:2017dux}
  S.~Chigusa, T.~Moroi and Y.~Shoji,
  Phys.\ Rev.\ Lett.\  {\bf 119} (2017) no.21,  211801
  [arXiv:1707.09301 [hep-ph]].

\bibitem{Chigusa:2018uuj}
  S.~Chigusa, T.~Moroi and Y.~Shoji,
  Phys.\ Rev.\ D {\bf 97} (2018) no.11,  116012
  [arXiv:1803.03902 [hep-ph]].

\bibitem{Andreassen:2017rzq}
  A.~Andreassen, W.~Frost and M.~D.~Schwartz,
  Phys.\ Rev.\ D {\bf 97} (2018) no.5,  056006
  [arXiv:1707.08124 [hep-ph]].

\bibitem{Coleman:1977py}
  S.~R.~Coleman,
  Phys.\ Rev.\ D {\bf 15} (1977) 2929;
  Erratum: [Phys.\ Rev.\ D {\bf 16} (1977) 1248].

\bibitem{Callan:1977pt}
  C.~G.~Callan, Jr. and S.~R.~Coleman,
  Phys.\ Rev.\ D {\bf 16} (1977) 1762.

\bibitem{Coleman:aspectsof}
  S.~Coleman, ``Aspects of Symmetry,''
  Cambridge University Press (1985) 265.


\bibitem{Claudson:1983et}
  M.~Claudson, L.~J.~Hall and I.~Hinchliffe,
  Nucl.\ Phys.\ B {\bf 228} (1983) 501.

\bibitem{Kusenko:1995jv}
  A.~Kusenko,
  Phys.\ Lett.\ B {\bf 358} (1995) 51
  [hep-ph/9504418].

\bibitem{Kusenko:1996jn}
  A.~Kusenko, P.~Langacker and G.~Segre,
  Phys.\ Rev.\ D {\bf 54} (1996) 5824
  [hep-ph/9602414].

\bibitem{Dasgupta:1996qu}
  I.~Dasgupta,
  Phys.\ Lett.\ B {\bf 394} (1997) 116
  [hep-ph/9610403].

\bibitem{Moreno:1998bq}
  J.~M.~Moreno, M.~Quiros and M.~Seco,
  Nucl.\ Phys.\ B {\bf 526} (1998) 489
  [hep-ph/9801272].

\bibitem{John:1998ip}
  P.~John,
  Phys.\ Lett.\ B {\bf 452} (1999) 221
  [hep-ph/9810499].

\bibitem{Cline:1998rc}
  J.~M.~Cline, J.~R.~Espinosa, G.~D.~Moore and A.~Riotto,
  Phys.\ Rev.\ D {\bf 59} (1999) 065014
  [hep-ph/9810261].

\bibitem{Cline:1999wi}
  J.~M.~Cline, G.~D.~Moore and G.~Servant,
  Phys.\ Rev.\ D {\bf 60} (1999) 105035
  [hep-ph/9902220].

\bibitem{Konstandin:2006nd}
  T.~Konstandin and S.~J.~Huber,
  JCAP {\bf 0606} (2006) 021
  [hep-ph/0603081].

\bibitem{Wainwright:2011kj}
  C.~L.~Wainwright,
  Comput.\ Phys.\ Commun.\  {\bf 183} (2012) 2006
  [arXiv:1109.4189 [hep-ph]].

\bibitem{Akula:2016gpl}
  S.~Akula, C.~Balázs and G.~A.~White,
  Eur.\ Phys.\ J.\ C {\bf 76} (2016) no.12,  681
  [arXiv:1608.00008 [hep-ph]].

\bibitem{Masoumi:2016wot}
  A.~Masoumi, K.~D.~Olum and B.~Shlaer,
  JCAP {\bf 1701} (2017) no.01,  051
  [arXiv:1610.06594 [gr-qc]].

\bibitem{Espinosa:2018hue}
  J.~R.~Espinosa,
  JCAP {\bf 1807} (2018) no.07,  036
  [arXiv:1805.03680 [hep-th]].

\bibitem{Jinno:2018dek}
  R.~Jinno,
  arXiv:1805.12153 [hep-th].

\bibitem{Espinosa:2018szu}
  J.~R.~Espinosa and T.~Konstandin,
  JCAP {\bf 1901} (2019) no.01,  051
  [arXiv:1811.09185 [hep-th]].

\bibitem{Athron:2019nbd}
  P.~Athron, C.~Balázs, M.~Bardsley, A.~Fowlie, D.~Harries and G.~White,
  arXiv:1901.03714 [hep-ph].


\bibitem{Coleman:1977th}
  S.~R.~Coleman, V.~Glaser and A.~Martin,
  Commun.\ Math.\ Phys.\  {\bf 58} (1978) 211.

\bibitem{Blum:2016ipp}
  K.~Blum, M.~Honda, R.~Sato, M.~Takimoto and K.~Tobioka,
  JHEP {\bf 1705} (2017) 109
  Erratum: [JHEP {\bf 1706} (2017) 060]
  [arXiv:1611.04570 [hep-th]].

\end{thebibliography}
\end{document}